\documentclass[]{article}
\usepackage{PRIMEarxiv}

\usepackage[utf8]{inputenc} 
\usepackage[T1]{fontenc}    
\usepackage{hyperref}       
\usepackage{url}            
\usepackage{booktabs}       
\usepackage{amsfonts}       
\usepackage{nicefrac}       
\usepackage{microtype}      
\usepackage{lipsum}
\usepackage{fancyhdr}       
\usepackage{graphicx}       
\usepackage{multicol}

\usepackage[
backend=biber,
style=ieee,
citestyle=numeric-comp
]{biblatex}

\graphicspath{{media/}}     

\title{Model Share AI: An Integrated Toolkit for Collaborative Machine Learning Model Development, Provenance Tracking, and Deployment in Python
}

\author{
  Heinrich Peters\\
  Columbia University, Modelshare AI \\
  New York\\
  \texttt{\ hp2500@columbia.edu} \\
   \And
  Michael Parrott \\
  Columbia University, Modelshare AI  \\
  New York\\
  \texttt{\ mp3675@columbia.edu} \\
}

\begin{document}
\maketitle

\begin{abstract}
Machine learning (ML) has the potential to revolutionize a wide range of research areas and industries, but many ML projects never progress past the proof-of-concept stage. To address this issue, we introduce Model Share AI (AIMS), an easy-to-use MLOps platform designed to streamline collaborative model development, model provenance tracking, and model deployment, as well as a host of other functions aiming to maximize the real-world impact of ML research. 
AIMS features collaborative project spaces and a standardized model evaluation process that ranks model submissions based on their performance on unseen evaluation data, enabling collaborative model development and crowd-sourcing. Model performance and various model metadata are automatically captured to facilitate provenance tracking and allow users to learn from and build on previous submissions. Additionally, AIMS allows users to deploy ML models built in Scikit-Learn, TensorFlow Keras, PyTorch, and ONNX into live REST APIs and automatically generated web apps with minimal code. The ability to deploy models with minimal effort and to make them accessible to non-technical end-users through web apps has the potential to make ML research more applicable to real-world challenges.

\end{abstract}


\section{Introduction}

Machine learning (ML) has the potential to revolutionize a wide range of research areas and industries, providing data-driven solutions to important societal problems. However, researchers and practitioners lack easy-to-use, structured pathways to collaboratively develop and rapidly deploy ML models. Traditionally, researchers have been using version-control systems like GitHub in combination with custom model evaluation and benchmarking experiments to ensure reproducibility and to compare models. However, larger-scale collaboration and crowd-sourcing are severely limited in the absence of standardized tasks and standardized processes for model sharing and evaluation. Additionally, most models developed by data scientists do not progress past the proof-of-concept stage and are never deployed \cite{davenport_is_2022, siegel_models_2022}, preventing the wider audience from participating in the promise of applied ML research. While the recent rise of platforms and tools like Hugging Face Hub \cite{noauthor_hugging_2023}, TensorFlow Hub \cite{noauthor_tensorflow_2023}, and ML Flow \cite{chen_developments_2020, noauthor_mlflow_2023, zaharia_accelerating_2018}, illustrates the demand for open-source model repositories and MLOps solutions, barriers of entry are still high for researchers, educators, and practitioners from non-technical disciplines. Model Share AI (AIMS) addresses this problem by providing a lightweight, easy-to-use alternative. In a few lines of code, users can create Model Playgrounds - standardized ML project spaces that offer an all-in-one MLOps toolkit for collaborative model improvement, experiment tracking, model metadata analytics, and instant model deployment, allowing researchers and other data scientists to rapidly share, improve, and learn from ML models in one streamlined workflow.

\section{Related Work}
The primary objective of AIMS is to offer an easy pathway to organize ML projects by reducing common and complex tasks down to minimal code. Its approach builds on the work of various projects focused on providing value for common ML tasks (e.g., deployment, model improvement, version tracking, etc.). Currently, there are several open-source tools and platforms, such as Hugging Face Hub \cite{noauthor_hugging_2023}, TensorFlow Hub \cite{noauthor_tensorflow_2023}, ML Flow \cite{chen_developments_2020, noauthor_mlflow_2023, zaharia_accelerating_2018}, and OpenML \cite{feurer_openml-python_2021, vanschoren_openml_2014, van_rijn_openml_2013}, providing ML model repositories where researchers can find and download model objects or deploy models. Hugging Face Hub \cite{noauthor_hugging_2023} is a platform allowing users to share pre-trained models, datasets, and demos of ML projects. It has GitHub-inspired features for code-sharing and collaboration, such as discussions and pull requests, and it includes a pathway for model deployment into API endpoints using AWS SageMaker. Similarly, TensorFlow Hub \cite{noauthor_tensorflow_2023} is a repository and library for reusable ML in TensorFlow, enabling users to fine-tune and deploy deep learning models. Deployment is facilitated by TensorFlow Serving \cite{olston_tensorflow-serving_2017}, which allows users to make their models accessible on a server through API endpoints. MLFlow \cite{chen_developments_2020, noauthor_mlflow_2023, zaharia_accelerating_2018} is an open-source platform that manages the end-to-end ML lifecycle. It provides experiment tracking, code packaging, a model registry, model serving, and integration with all popular ML frameworks. While Hugging Face Hub, TensorFlow Hub, and MLFlow are well suited for large-scale deep learning tasks, OpenML \cite{feurer_openml-python_2021, vanschoren_openml_2014, van_rijn_openml_2013}, focuses more on classic ML and model reproducibility. Here, researchers can share, explore, and experiment with ML models, datasets, and workflows, but there is currently no option for model deployment. The OpenML API provides access through various programming languages, but users can also employ an OpenML web interface to browse and visualize data, models, and experiments.
Despite these important developments, there is a lack of open-source technologies allowing users a simple, centralized way to find, analyze, download, and collaboratively improve trained ML models supported by rich model metadata analytics, and easily share models from their local environment directly into live REST APIs and prebuilt web applications in a single tightly integrated workflow. With a strong focus on ease of use, AIMS provides a simple, yet versatile approach to these challenges.

\section{Model Share AI}
AIMS provides standardized ML project spaces (Model Playgrounds) with accessible MLOps features designed to support collaborative model development, model metadata analytics, and model deployment, as well as a host of other functions aiming to maximize the real-world impact of ML research. As such, it simplifies, combines, and extends the capabilities of current solutions in 4 important ways:

\begin{enumerate}
    \item Collaborative model development: AIMS facilitates collaborative model development and crowd-sourcing through standardized model evaluation procedures, experiment tracking, and competitions. 
    \item Model registry: AIMS hosts model objects, as well as model metadata, automatically extracted from each model submission, enabling users to analyze which modeling decisions lead to high performance on specific tasks.
    \item Model deployment: Model deployment is simplified considerably, allowing users to share their models into live REST APIs and pre-built web apps directly from their Python training environments with just a few lines of code.
    \item AIMS provides a wide range of supporting functionalities, including workflows for reproducibility, data sharing, code sharing, creating ML project portfolios, and more.
\end{enumerate}

\begin{figure}
  \centering
  \includegraphics[width=1\textwidth]{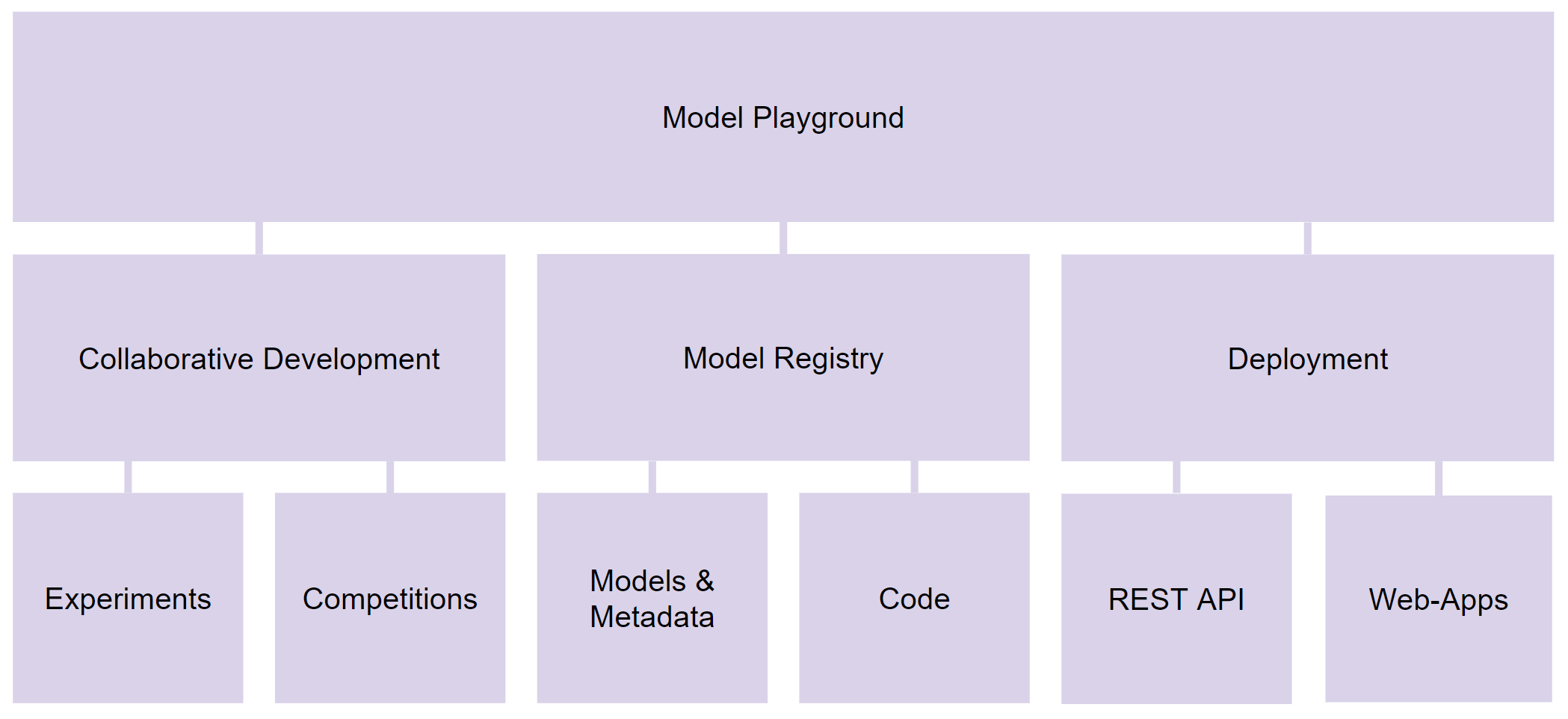}
  \caption{A Model Playground is a standardized ML project space, representing a ML task that is associated with a specific dataset and a task type (classification, regression). A Model Playground includes Experiments and Competitions for collaborative model development, a model registry with model metadata and user-generated code, as well as deployment functionalities including a live REST API and pre-built web-apps.}
  \label{fig:fig_overview}
\end{figure}

\subsection{Key Functions}
\paragraph{Collaborative model development.} A key feature of AIMS is its focus on collaborative model development and crowd-sourced model improvement, enabling teams to iterate quickly by allowing collaborators to build on each other's progress, even across libraries. For supervised learning tasks, users can collaboratively submit models into Experiments or Competitions associated with a Model Playground project in order to track model performance and rank submissions in standardized leaderboards. Experiments and Competitions are set up by providing evaluation data against which the predictions of submitted models are evaluated. Standardized model evaluations allow collaborators to track the generalized performance of their models along with a wide range of model metadata that are automatically extracted from submitted models and added to the model registry (see section below). Out of the box, AIMS calculates accuracy, f1-score, precision, and recall for classification tasks, and mean squared error, root mean squared error, mean absolute error, and $R^{2}$-scores for regression tasks, but users can submit custom evaluation functions for more flexibility. The main difference between Experiments and Competitions is that a proportion of the evaluation data is kept secret for Competitions, preventing participants from deliberately overfitting on evaluation data. Being able to submit models into shared Experiments enables ML teams to standardize tasks, rigorously track their progress, and build on each other's success, while Competitions facilitate crowd-sourced solutions for any ML task. Both Experiments and Competitions can be either public (any AIMS user can submit) or private (only designated team members can submit). Users can deploy any model from an Experiment or Competition into the REST API associated with their Model Playground with a single line of code. 

\paragraph{Model Registry.} Model versions are made available for each Model Playground and comprehensive model metadata are automatically extracted for each submitted model. In addition to evaluation metrics, this includes all hyperparameter settings for Scikit-Learn models and model architecture data (such as layer types and dimensions, number of parameters, optimizers, loss function, memory size) for Keras and Pytorch models. Users can also submit any additional metadata they choose to capture. Model metadata are integrated into Competition and Experiment leaderboards, enabling users to analyze which types of models tend to perform well for a specific ML task. Users can either visually explore leaderboards on their Model Playground page or they can download leaderboards into Pandas data frames to run their own analyses. There is also a set of AIMS methods designed to visualize model metadata. For example, models can be compared in a color-coded layout showing differences in model architectures and hyperparameter settings. Furthermore, users can instantiate models from the AIMS model registry into reproducible environments in a single line of code. They can either instantiate untrained model objects with chosen hyperparameter settings and architecture if they want to retrain the model or use it for a different task, or they can instantiate trained models, including parameter values. Taken together, these functions are designed to streamline the management, collaboration, and deployment of ML models, enhancing their discoverability, reproducibility, and traceability throughout their lifecycle.

\paragraph{Instant model deployment.} AIMS currently allows users to deploy ML models built in Scikit-Learn \cite{pedregosa_scikit-learn:_2011}, Tensorflow Keras \cite{abadi_tensorflow_2016, chollet_keras_2015}, and Pytorch \cite{paszke_pytorch_2019} into live rest APIs rapidly with minimal code. Additionally, users can deploy models from any other ML framework by transforming them into ONNX (Open Neural Network Exchange) format - an open-source format for standardized representations of ML models with the goal of making them interoperable across tools and platforms. Each deployed model is associated with a Model Playground page on the AIMS website and a REST API endpoint hosted in a serverless AWS backend. End-users can either manually upload data to make predictions using an automatically generated web app on the Model Playground Page, or they can programmatically query the REST API associated with the model. In addition to auto-generated web apps, AIMS enables users to submit their own Streamlit apps. Out of the box, AIMS supports models built on tabular, text, image, audio, and video data. Allowing users to deploy models with minimal effort and making those models accessible to even non-technical end-users through web apps holds the promise of making ML research applicable to real-world challenges.

\begin{figure}
  \centering
  \includegraphics[width=1\textwidth]{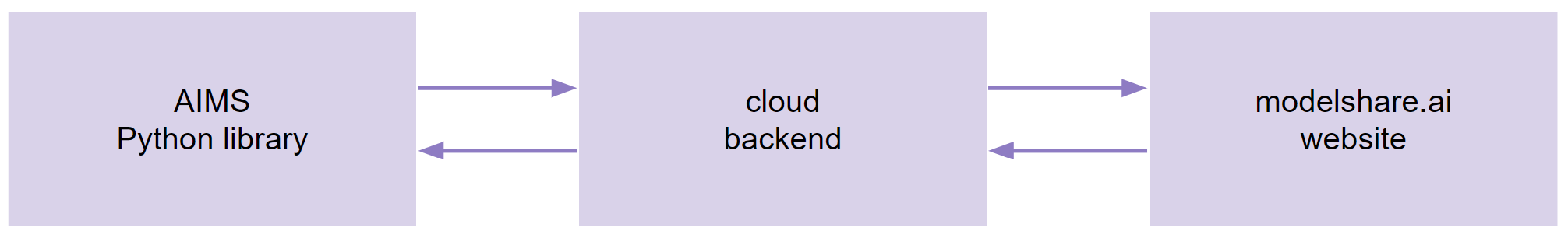}
  \caption{Overview of the AIMS architecture. The AIMS Python library allows users to create Model Playground pages, submit and deploy models, and analyze model metadata. The modelshare.ai website provides a graphical user interface to explore model metadata and generate predictions via auto-generated web apps. All required resources are automatically generated in scalable serverless cloud infrastructure.}
  \label{fig:fig_architecture}
\end{figure}

\subsection{Architecture}
AIMS consists of three main components: an open-source Python library, cloud backend resources, and the modelshare.ai website. The AIMS Python library (\href{https://github.com/AIModelShare/aimodelshare}{https://github.com/AIModelShare/aimodelshare}) is the main interface allowing users to set up Model Playground pages (including Experiments and Competitions), submit and deploy models, analyze model metadata, and reproduce model artifacts. It provides an accessible layer that facilitates the creation of the cloud backend resources that power REST APIs, as well as model evaluations and model metadata extraction. The ModelPlayground() class acts as a local representation of a Model Playground page and its associated REST API. It provides a range of methods to configure, change, and query Model Playground resources. A detailed overview of the Python library is provided below (3.3 AIMS workflow). 

The cloud backend hosts model objects and associated artifacts in S3 storage, while REST APIs are deployed into serverless lambda functions. Lambda functions are programs or scripts that run on high-availability AWS compute infrastructure. They can be invoked by various event sources (e.g., API calls) and scale automatically based on the volume of incoming requests. This means that users do not have to manage, maintain, or patch servers, and they only pay for the time and resources actually consumed, but not for idle time. The AIMS Python library allows users to automatically generate and deploy lambda functions based on specific properties of their ML tasks, without the need to explicitly manage any AWS resources. The most important lambda functions in the context of AIMS are the Evaluation Lambda, which computes evaluation metrics and extracts model metadata from submitted models, and the Main Lambda, which computes predictions on data submitted through the REST API. Runtime models are automatically packaged into Docker containers that run on lambda. Additionally, certain metadata are stored in a centralized Redis database that powers the modelshare.ai website. 

The AIMS website (\href{https://www.modelshare.ai}{https://www.modelshare.ai}) \cite{noauthor_model_nodate} hosts user profile pages, model pages, web apps, example code, and an official documentation page, as well as user-generated code and documentation for their specific ML projects (see SI \ref{app:tutorials}-\ref{app:model}).

\subsection{AIMS Workflow}
The AIMS workflow is designed to help teams collaboratively and continuously train, evaluate, improve, select, and deploy models using standardized ML project spaces or Model Playgrounds. After training a model, users submit their model to a Competition or Experiment associated with a Model Playground. The model is then automatically evaluated, and model metadata are extracted. Evaluations and metadata are made available via a graphical user interface on the Model Playground page or can be queried using the AIMS Python library, enabling users to analyze which types of models perform well on a given learning task. This information can then be used to either improve the training process and submit more models or to select a model and instantly deploy it with a single line of code. Deployed models can easily be swapped out if they degrade over time or if they are outperformed by new submissions.

\begin{figure}
  \centering
  \includegraphics[width=1\textwidth]{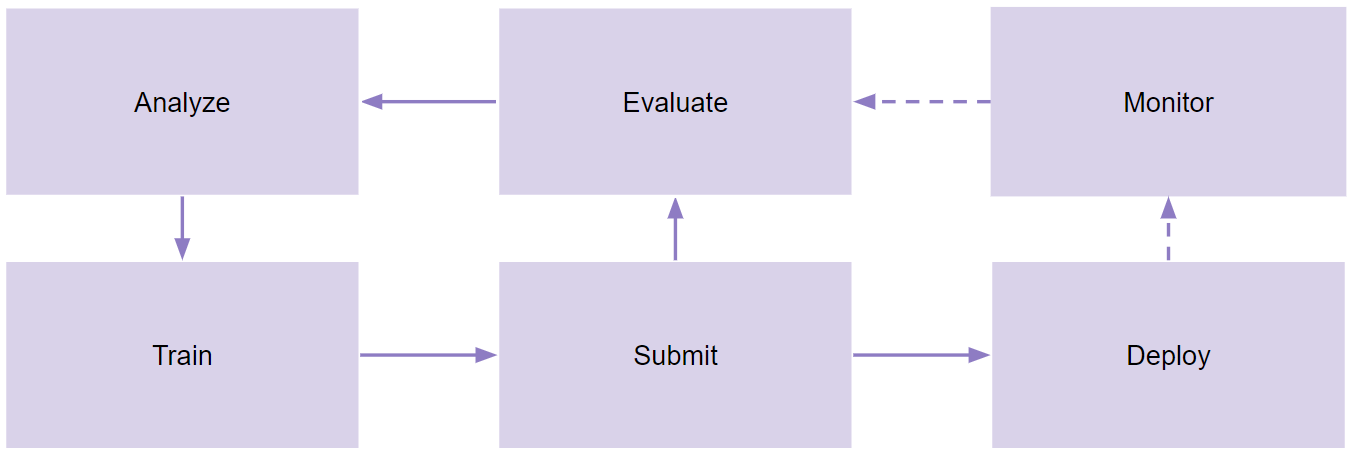}
  \caption{MLOps workflow with AIMS. Users iteratively train, submit, evaluate, and analyze their models. As contributors have access to model evaluations, model architecture metadata, and reproducible model objects of previous submissions, they can rapidly develop high-performing models. Submitted models can easily be deployed into live REST APIs. Deployed runtime models can be monitored and seamlessly be swapped out against newly submitted models.}
  \label{fig:fig_aims_workflow}
\end{figure}

In order to use AIMS, users first need to create an AIMS user profile and generate credentials through the AIMS website (\href{https://www.modelshare.ai}{https://www.modelshare.ai}) \cite{noauthor_model_nodate}. Additionally, users are required to generate AWS credentials if they wish to deploy models into their own AWS resources. 

The AIMS workflow is centered around the concept of Model Playgrounds. A Model Playground is a standardized ML project space, representing a ML project that is associated with a specific dataset and a task type, such as classification or regression. A Model Playground is first instantiated locally using the ModelPlayground() class of the AIMS Python library. Here, users need to specify the input\_type (tabular, text, image, audio, etc.) and the task\_type (classification, regression), as well as select whether their Model Playground should be public or private. Private Model Playgrounds can only be accessed by invited collaborators, whereas public Model Playgrounds are open to all AIMS users. After instantiating their Model Playground object, users can create an online Model Playground page by calling the create() method and submitting their evaluation data. This generates a fully functioning Model Playground Page on the AIMS website, including a placeholder REST API, and enables users to submit models into associated Experiments and Competitions. 

Once a Model Playground Page is created, users can start submitting models to Experiments or Competitions and deploying models into the Model Playground’s REST API. A model submission includes a model object (Scikit-Learn, Keras, PyTorch, ONNX), a preprocessor function, and a set of predicted values corresponding to the previously submitted evaluation data. The predictions are then evaluated against the previously submitted evaluation data, and model metadata are automatically extracted from model objects. Users can submit additional metadata in dictionary format using the custom\_metadata argument of the submit\_model() method. After submitting one or more models, users can explore evaluations and model metadata on the associated Model Playground page or query this information for further analysis using the get\_leaderboard() and compare\_models() methods. To inspect and improve model objects, users can instantiate models from the leaderboard using the instantiate\_model() method. Alternatively, users can instantly deploy a model using the deploy\_model() method by referring to the model’s leaderboard version number. Additionally, users should submit example data, which will help end-users to format their input data correctly, and y training data to help the web app and prediction API transform raw model outputs into the correct labels. As mentioned above, the process does not stop when a model is deployed. Users can submit and analyze more models, informed by previous submissions, and easily monitor and swap out the runtime model using the update\_runtime\_model() method. Detailed tutorials and documentation can be found in SI \ref{app:tutorials}.

\begin{figure}
  \centering
  \includegraphics[width=1\textwidth]{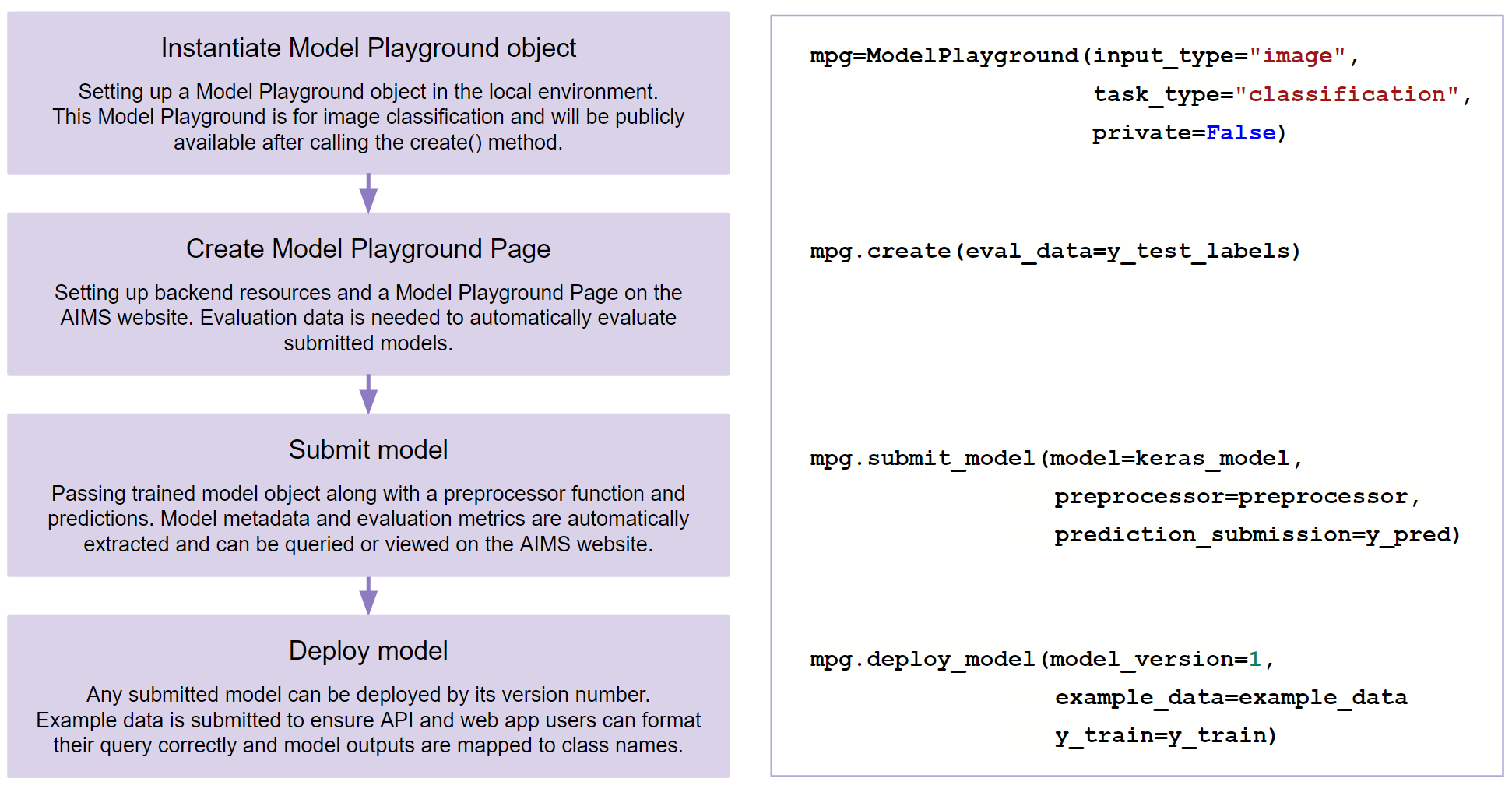}
  \caption{Model deployment process: Models can be deployed with just a few lines of code. After instantiating a local Model Playground object, users can create a Model Playground page that is ready for model submissions. Submitted models are automatically evaluated and can be deployed into live REST APIs.}
  \label{fig:fig_code_example}
\end{figure}

\section{Impact}
Collaborative model improvement and crowd-sourcing are important, not only because they make teams more efficient but also because they foster diverse perspectives. These collective efforts enabled by AIMS contribute to the democratization of AI, providing access to resources and knowledge for a wider audience, enabling community-driven innovation, and accelerating progress in the field. For example, the AIMS Competition feature is uniquely positioned to facilitate crowd-sourced research projects in various disciplines utilizing the common task method \cite{salganik_measuring_2020}. 
The AIMS model registry promotes discoverability,  transparency, and reproducibility by providing a centralized platform for researchers, developers, and data scientists to find and access pre-trained models and their associated metadata. Sharing model metadata, such as model architectures, hyperparameters, training data, and evaluation metrics, allows researchers to verify previous results and build upon each other's work. The repository also acts as a valuable educational resource, offering students, educators, and self-learners the opportunity to study various ML models, techniques, and best practices.
With its focus on ease of use, AIMS is uniquely positioned to realize several important missions with regard to model deployment. First, simplifying model deployment is essential because it lowers barriers to entry, allowing more developers, researchers, and organizations to incorporate ML in their projects, regardless of their level of expertise. As deployment becomes more accessible, it fosters a broader adoption of ML solutions across diverse industries, leading to innovative applications and enhancements in products and services. Second, easier deployment also translates to cost-effectiveness, saving resources and time needed for implementation, such that developers can dedicate more time to model training, tuning, and evaluation. Third, by making ML deployment easier, AIMS contributes to the democratization of AI, providing smaller businesses and individuals the opportunity to utilize ML technologies. Ultimately, streamlining the deployment process encourages experimentation and innovation in ML, allowing researchers and developers to iterate faster and explore novel ideas and techniques, leading to improved research, products, and services. 

\section{Future Work}
While AIMS provides a greatly simplified approach to collaborative model development and deployment, there are opportunities for further improvements. Firstly, it is important to strike the right balance between flexibility and ease of use. A highly flexible platform may allow researchers to use various ML frameworks and libraries, accommodate a wide range of data sources and formats, and support custom deployment strategies. However, this flexibility may come at the cost of increased complexity and a harder learning curve for users. On the other hand, a more user-friendly solution might provide a simpler, more streamlined interface with pre-built templates and workflows but may lack the flexibility to accommodate unique or complex requirements. By default, AIMS prioritizes standardization and ease of use, making it attractive for researchers, educators, and data scientists who are interested in quick and lightweight solutions. For users who want more flexibility, AIMS provides customizations, such as custom AWS lambda functions, custom containers, custom model evaluation functions, and custom metadata submissions. We will continue to extend the functionality of AIMS to make it useful for a wide range of users and applications. This includes making the platform compatible with more advanced model types, task types, and ML frameworks. Relatedly, future work will include additional MLOps functionality, including improved pathways for monitoring, model explainability, continuous retraining, and automated ML (AutoML) \cite{hutter_automated_2019, he_automl_2021}. The latter point is of special interest, as each new model submission contributes to a growing repository of evaluated, reusable ML models, including rich model metadata. These resources can be utilized to suggest pre-trained models that are expected to work well for a given task or enable users to run their own analyses to choose pre-trained models. We expect AIMS to become a valuable resource for researchers interested in meta-learning \cite{hospedales_meta-learning_2020, finn_model-agnostic_2017} and related problems. Additionally, we expect AIMS to be used increasingly by researchers from various disciplines, including social sciences and natural sciences, trying to solve substantive questions through ML. We are planning to further accommodate their diverse needs in order to promote cross-fertilization and widespread participation in the promise of applied ML research. 

\section{Conclusion}
AIMS provides a versatile yet easy-to-use approach to collaborative model development, model metadata analytics, and model deployment. AIMS enables users to quickly find, analyze, download, and collaboratively improve trained ML models, and to share models from their local environment directly into live REST APIs and pre-built web applications in a single tightly integrated workflow. Compared to existing solutions, AIMS significantly lowers the barriers of entry to ML research, making it attractive to a large group of researchers, educators, and data scientists. We are confident that AIMS will become a widely used tool and maximize the real-world impact of  ML research.

\newpage
\section*{Acknowledgments}
We are profoundly grateful to our early and repeated organizational backers. Columbia’s QMSS program and Director Greg Eirich were early supporters. Professor David Park, Columbia University’s former GSAS Dean of Strategic Initiatives, helped to carve out a strategy that led to the project’s eventual fundraising success. A special acknowledgment goes to Columbia University's ISERP Startup Center. Their funding enabled Dr. Michael Parrott, our founding Director, to assemble an early team and develop a prototype of the platform. This early funding and technical prototype laid the groundwork for further large-scale funding and support from the Alfred P. Sloan Foundation. Director Josh Greenburg and the Tech program team at Sloan gave us the momentum and repeated support to innovate and build what has become the first research and education-facing MLOps platform. Without their grant resources and their belief in our success, this project would have been impossible. Of course, resources are only a starting point. We must highlight the exceptional contributions of Sam Alsmadi, who developed the AIMS website, and Gretchen Street, who led the way on library user security, ML datasets, and documentation. Finally, we want to thank the talented alumni who have contributed code over the years. This platform would not have been possible without all your hard work and innovation!

\section*{Author Contributions}
M.P. developed the original idea for the Model Share AI platform in 2019, assembled the team, and led the way to the platform's completion as the chief engineer and team lead. (CRediT: Conceptualization, Methodology, Software, Resources, Writing - Review \& Editing, Supervision, Project administration, Funding acquisition)
H.P. has taken a leading role in the development of the AIMS Python library and has had pivotal impact on ideation and product development in his role as lead ML engineer since 2020. (CRediT: Conceptualization, Methodology, Software, Writing - Original Draft)

\newpage
\printbibliography
\newpage
\section*{Supplementary Information}
\appendix

\section{Tutorials, Code, and Documentation}
\label{app:tutorials}

\begin{itemize}

\item 
\href{https://colab.research.google.com/github/AIModelShare/aimodelshare_tutorials/blob/main/modelshareai/Text_Classification_QuickStart_IMDB.ipynb}{Text Classification Tutorial}

\item 
\href{https://colab.research.google.com/github/AIModelShare/aimodelshare_tutorials/blob/main/modelshareai/QST_Tabular_Classification.ipynb}{Tabular Classification Tutorial}

\item 
\href{https://colab.research.google.com/github/AIModelShare/aimodelshare_tutorials/blob/main/modelshareai/Keras_Image_Classification_Quick_Start_Tutorial.ipynb}{Image Classification Tutorial}

\item 
\href{https://colab.research.google.com/github/AIModelShare/aimodelshare_tutorials/blob/main/modelshareai/Keras_Video_Classification_Quick_Start_Tutorial(1).ipynb}{Video Classification Tutorial}

\item 
\href{https://www.modelshare.ai/notebooks/notebook:229}{Regression Tutorial}

\item 
\href{https://github.com/AIModelShare/aimodelshare}{Official Github Page}

\item 
\href{https://aimodelshare.readthedocs.io/en/latest/index.html}{Official Documentation}

\item 
\href{https://www.modelshare.ai/}{Modelshare AI Website}

\end{itemize}

\newpage
\section{Example of User Profile Page}
\label{app:profile_page}

\begin{figure}[h!]
  \centering
  \includegraphics[width=1\textwidth]{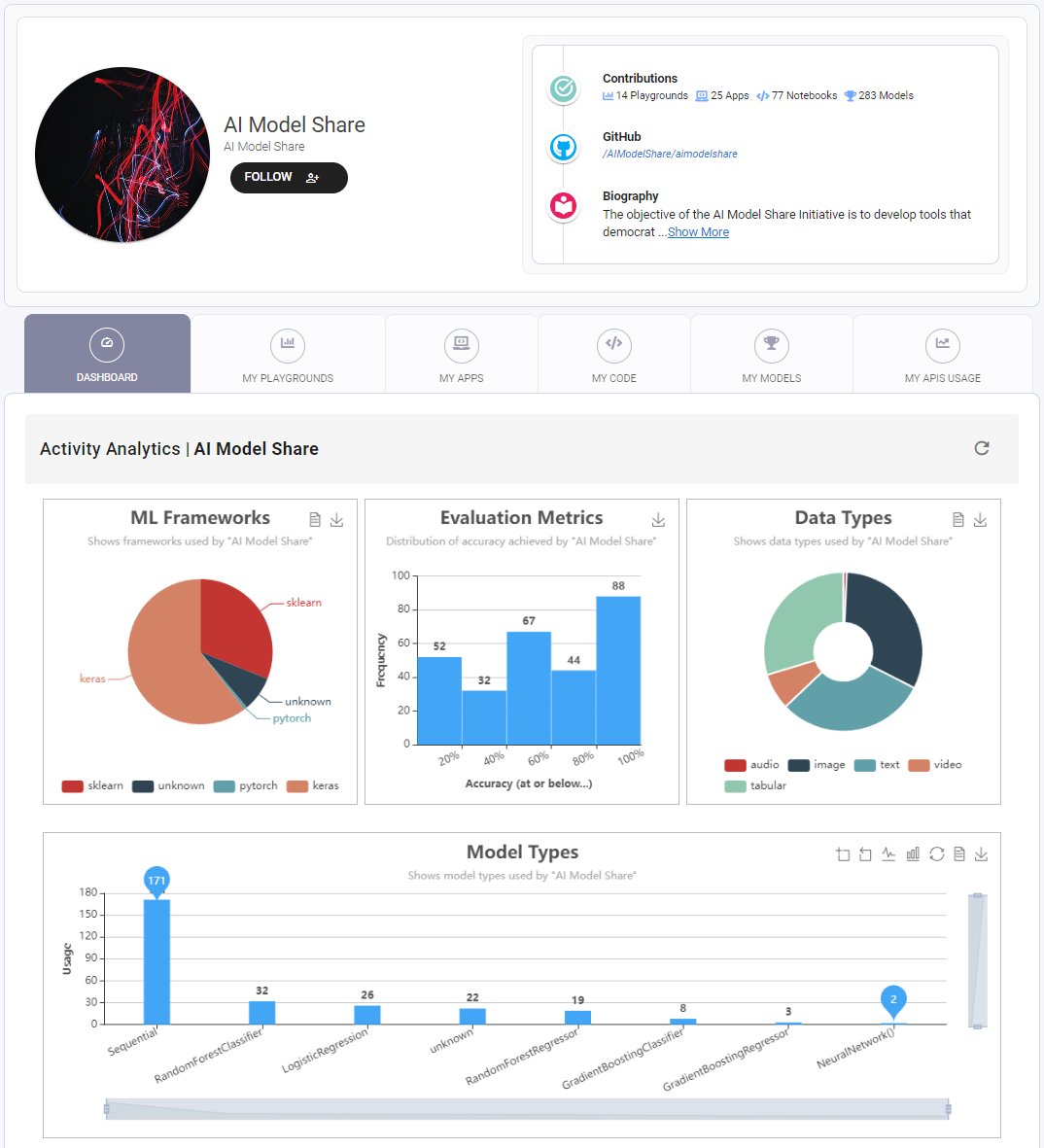}
  \caption{Example of a user profile page. A User Profile Page shows aggregate information about a user's activity (Dashboard), and provides access to Model Playgrounds, apps, code, and models created by the user.}
\end{figure}

\newpage
\section{Example of Model Playground Page}
\label{app:playground}

\begin{figure}[h!]
  \centering
  \includegraphics[width=1\textwidth]{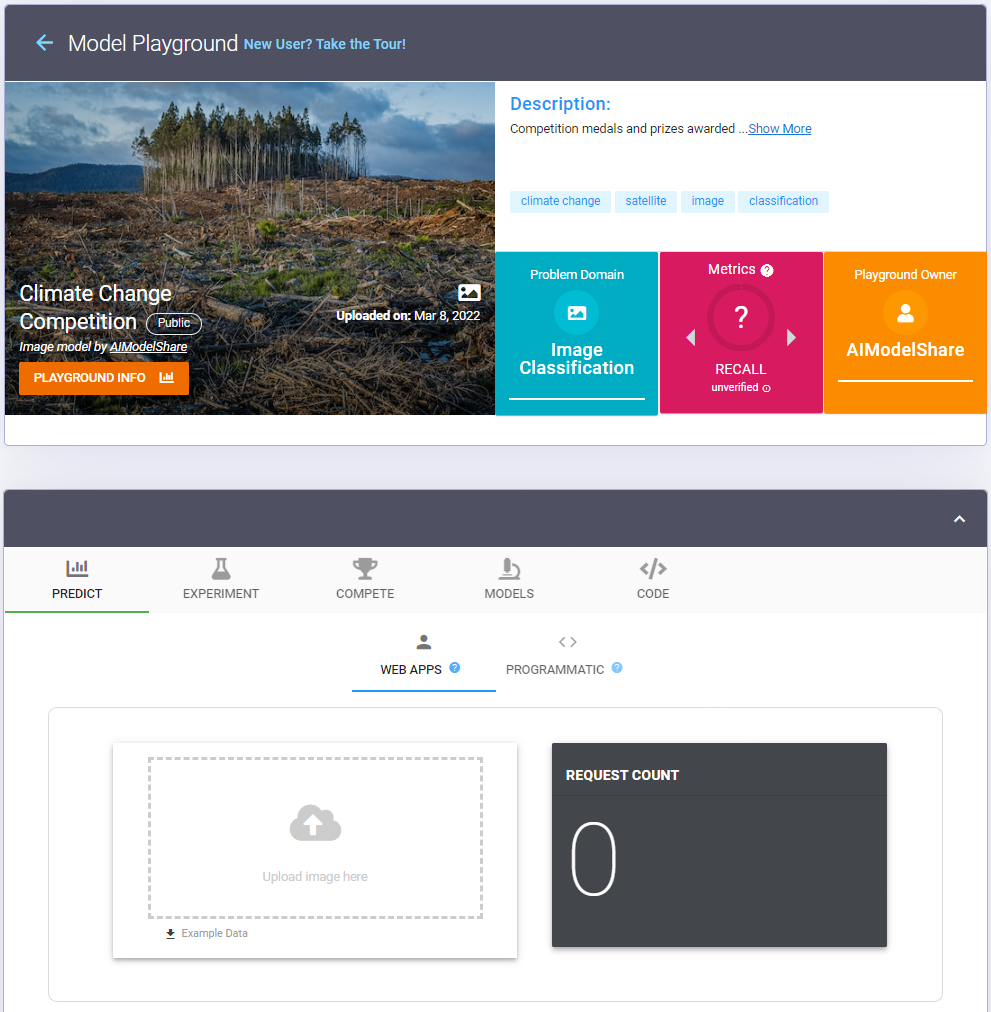}
  \caption{Example of a Model Playground page. The model playground page allows users to make predictions by uploading data or accessing rebuilt code to make API calls, access Experiment and Competition leaderboards, inspect and compare model architectures, and access code submissions.}
\end{figure}

\newpage
\section{Example of Competition Leaderboard}
\label{app:competition}

\begin{figure}[h!]
  \centering
  \includegraphics[width=1\textwidth]{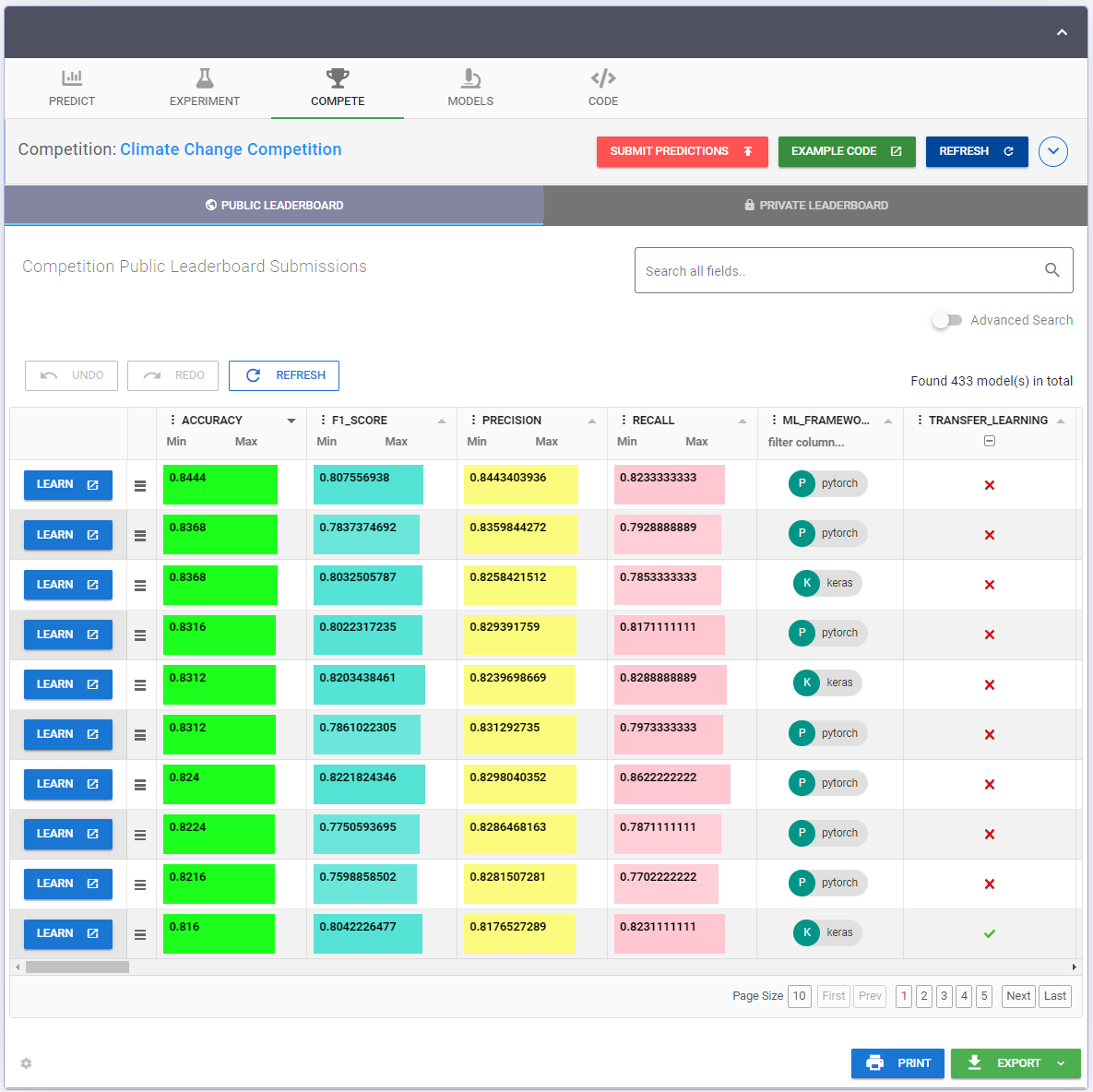}
  \caption{Example of a Competition Leaderboard. Models submitted to Experiments or Competitions are automatically evaluated and can be ranked according to their evaluation metrics. Model metadata, such as hyperparameter settings, layer types, and dimensions, are depicted in the table (not visible in the screenshot). In order to inspect individual models, users can click on "Learn", which will take them to a Model Details page.}
\end{figure}

\newpage
\section{Example of Model Details Page}
\label{app:model}

\begin{figure}[h!]
  \centering
  \includegraphics[width=1\textwidth]{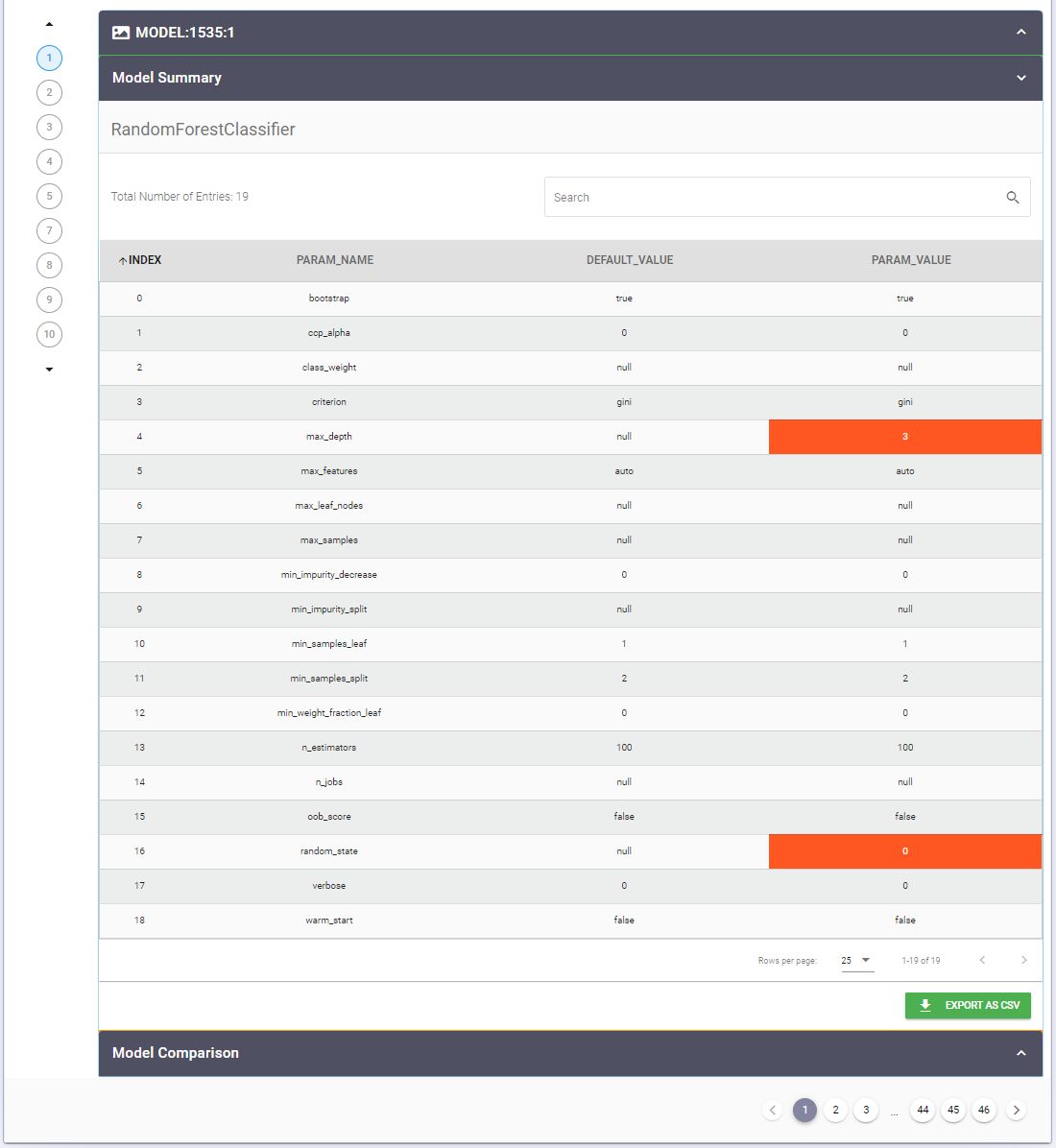}
  \caption{Example of a model and its hyperparameter settings. Users can inspect the hyperparameter settings and model architectures of any model submitted to an Experiment or a Competition.}
\end{figure}

\end{document}